# Effect of Gap Width on Turbulent Transition in Taylor-Couette Flow

Chang-quan Zhou[1, 2], Hua-Shu Dou[1*], Lin Niu[1], Wen-qian Xu[1, 3]

1. *Faculty of Mechanical Engineering, Zhejiang Sci-Tech University, Hangzhou 310018, China*
2. *School of Mechanical Engineering, Zhejiang University of Water Resources and Electric Power, Hangzhou 310018, China*
3. *School of Mechanical Engineering, Hangzhou Dianzi University, Hangzhou 310018, China*

**\*Corresponding author:** Dou, E-mail: huashudou@zstu.edu.cn

**Abstract** Simulations of the transitional flow in Taylor-Couette configuration are carried out to study the effect of the gap width on turbulent transition. The research results show that, under the same radius and the rotating speed of the inner cylinder, as the gap width increases, the flow becomes more stable. It is discovered that the average velocity distribution in the gap approaches the free vortex flow as the width increase and the stability of the flow is enhanced. It is found that, as the gap width increases, the maximum of the energy gradient function (from the energy gradient theory) in the gap decreases, which delays the turbulent transition. As such, the larger the gap width, the later the transition occurs. As the gap width increases, the Reynolds number based on the gap width alone is not able to characterize the flow behavior in Taylor-Couette flows, and the effect of the radius ratio should be taken into account.

**Kay words:** Taylor-Couette; Turbulent Transition; Singularity; Negative spike; Stability; Reynolds number

0. Introduction

Turbulence is one of the most difficult problems in classical physics, which has been studied for over 100 years. However, the physical mechanism of turbulent generation is still a challenge[1-5]. The Taylor-Couette flow between two coaxial cylinders is a typical canonical problem which has been extensively studied in the community[6-12]. However, most researches in Taylor-Couette flows are focused on the cases of narrow gaps. In recent years, there have been several studies on the situation of wide gaps between the two cylinders[13-19]. In these works, it is found that it is difficult to generate turbulence when the inner cylinder rotates and the outer cylinder is fixed[12-16]. For the Keplerian regime, the flow is nonlinear stable even at high Reynolds number, where turbulence cannot be sustained[12, 14, 16]. The physical reason for this phenomenon has not been well understood yet.

In these studies, the shear Reynolds number ($Re_h$) based on the gap width is often used as an important dimensionless parameter for characterizing the flow stability and the generation of turbulence[20]. In the following, we discuss the status where the inner cylinder is rotating and the outer cylinder is fixed. When the fluid kinematic viscosity is given, the Reynolds number $Re_h$ can be changed by varying the tangential velocity and the gap width. In terms of the Reynolds number, it is commonly considered that a larger $Re_h$ will lead to the flow more unstable. When the width of the gap between the two cylinders is fixed, increasing the rotating speed of the inner cylinder leads to a large Reynolds number, and the flow is more unstable[6-12]. However, when the rotating speed of the inner cylinder remains constant, a larger gap width gives a large Reynolds number $Re_h$. In this case, the stability of flows in the gap may vary inversely with the increase of the Reynolds number, which is opposite to the human intuition. If it is true, the flow stability in Taylor-Couette flow does not depend on the Reynolds number $Re_h$. In other words, the Reynolds number $Re_h$ does not reflect the stability of flows in Taylor-Couette devices. As such, the physical mechanism for causing the flow more stable with increasing $Re_h$ needs to be clarified.

The energy gradient theory has been proposed to study the flow stability and turbulent transition[5]. For both wall-bounded flow and free shear flow, it has obtained very good agreement with the experimental data and the computational results with direct numerical simulations (DNS) and large eddy simulations (LES). With this theory, turbulence is produced by the singularities of the Navier-Stokes equation, which cause discontinuities of the streamwise velocity. In turn, the velocity discontinuity transfers the energy from the main stream to the fluctuating turbulence.

In this study, numerical simulations with large eddy simulation (LES) for Taylor-Couette flow are carried out for three cases of gap widths. Effects of the Reynolds number $Re_h$ on flow stability and turbulence generation are investigated. It is found that the flow becomes more stable with the increase of $Re_h$ as the increase of the gap width and thus turbulent transition is delayed. The physical reason for this phenomenon is discovered that the flow tends towards free vortex flow as the gap width increases, while the free vortex flow is the most stable[5]. Finally, it is concluded that the energy gradient function, which stands for a local Reynolds number, decides the stability of flow.

1. Turbulence initiation in Taylor-Couette flows

In this study, the outer cylinder does not rotate. The diameter and the rotating speed of the inner cylinder are fixed and the gap width between the two cylinders is changed. The stability of the internal flow in the Taylor-Couette flow is studied to explore the influence of the Reynolds number based on the gap width. When the gap width increases, the Reynolds number increases. Then the flow characteristics in the gap for the three cases of gap widths are compared and further analyzed.

Now, two Reynolds numbers are defined as follows,

$$Re_{R1} = \frac{u_{\theta 1} R_1}{\nu},\qquad(1)$$

$$Re_h = \frac{u_{\theta 1} h}{\nu},\qquad(2)$$

where, $h = R_2 - R_1$ is the gap width, and $R_1$ and $R_2$ are the radii of the inner and outer cylinders, respectively, $u_{\theta 1}$ and $u_{\theta 2}$ are the tangential velocities of the inner and outer cylinders, respectively, and $\nu$ is the kinematic viscosity.

In Eqs.(1) and (2), $Re_{R1}$ is called the rotating Reynolds number, and $Re_h$ is called the shear Reynolds number[20]. This shear Reynolds number is widely used in most of the researches of Taylor-Couette flows[12-20]. It is well known that turbulent transition in the gap of the Taylor-Couette flow will occur as the increase of the Reynolds number $Re_{R1}$ [5-11]. In the present study, $Re_{R1}$ is fixed, and effect of $Re_h$ on initiation of turbulent transition is investigated by varying the width of the cylinder gap. It will be seen that the result of effect of $Re_h$ on turbulent transition is surprising. Although the effect of the radius ratio should be taken into account when the gap width is large, the mechanism to cause this change of flow behavior should be understood.

The angular velocities of the inner and outer cylinders are expressed by $\omega_1$ and $\omega_2$, respectively. Then, the velocity distribution in the Taylor-Couette flow for the laminar flow is expressed as follows which is obtained from the Navier-Stokes equation[5-6,11]

$$u = Ar + \frac{B}{r},\qquad(3)$$

where

$$A = \omega_1 \frac{(\eta^2 - \lambda)}{\eta^2 - 1} \text{ and } B = \omega_1 R_1^2 \frac{(1-\lambda)}{1-\eta^2}.\qquad(4)$$

In above equations, $\lambda = \omega_2 / \omega_1$, and $\eta = R_1 / R_2$.

As can be seen from Eqs. (3) and (4), the velocity distribution in the gap is composed of a forced vortex and a free vortex. As the gap width increases, the proportion of the free vortex increases, while the proportion of the forced vortex decreases. When the gap width is very large, the velocity distribution within the gap is close to that of the free vortex flow.

Dou showed that a free vortex is of a uniform energy field, and the free vortex flow is stable[5]. Any disturbance overlapped in the free vortex flow will be quickly damped and decay. In such a way, the larger the proportion of the free vortex, the more stable the flow is. Thus, turbulent transition will therefore be delayed in Taylor-Couette flows with wide gap[5, 21].

According to researches on flow stability and turbulent transition in Refs. [5, 21-24], the generation of turbulence depends on the occurrence of singularities in the Navier-Stokes equation, i.e. velocity discontinuities, where "negative spikes" appear in the temporal recording of the streamwise velocity.

It was suggested[10] that the instability in Taylor-Couette flows may be similar to other shear flows such as the Blasius boundary layer, where negative spike ("V" shaped) appears in the streamwise velocity. These results are supported by the direct numerical simulation (DNS) results in Berghout et al.[25].

The mechanism of flow stability in viscous flows has been explained by the energy gradient theory[5]. In the theory, the energy gradient function $K$, which is equivalent to a local Reynolds number, is defined as[5, 21]

$$K = \frac{\partial E/\partial n}{\partial H/\partial s} = \frac{\partial E/\partial n}{\partial E/\partial s + \phi/V} = \frac{V \partial E/\partial n}{V \partial E/\partial s + \phi} \qquad (5)$$

where $n$ is along the direction normal to the streamwise direction, $s$ is along the streamwise direction, $V$ is the total velocity, $\phi$ is the rate of energy dissipation, $H$ is the rate of loss of the total mechanical energy, and $E$ is the total mechanical energy for incompressible flow.

Dou proposed an Axiom that the theoretical velocity varies monotonously with the rate of loss of the total mechanical energy along the streamwise direction, $\partial H/\partial s$ [5]. If the value of $\partial H/\partial s$ is zero along a streamline, the velocity will theoretically be zero, which forms a singular point. At the singularity, "negative spike" will be produced in the streamwise velocity. The spike produced at the singular point will be the "power" to sustain the turbulent fluctuations.

For a given boundary condition, the magnitude of the velocity $u$ varies monotonically with the variation of $V \partial E/\partial s + \phi$ along a streamline, as shown in Eq. (5). Especially, when the value of $V \partial E/\partial s + \phi$ is zero, the velocity $u$ tends to zero theoretically. This is the singularity of the Navier-Stokes equation since the velocity is not differentiable at this location. Therefore, at such a singularity position, we obtain $u=0$ theoretically, and a spike is generated. This spike causes velocity fluctuations, and further, the flow eventually develops into turbulence at sufficient high Reynolds number. The "negative spike" in the streamwise velocity distribution is the origin of turbulence. This theory has obtained agreement with the numerical and experimental results for various flows such as pipe Poiseuille flow, plane Poiseuille flow, and plane Couette flow, and boundary layer flow, etc.[5, 21-24].

The principle of flow stability from the energy gradient theory is suitable both for Newtonian flow and non-Newtonian flow. The theory has obtained agreement with the experimental data in viscoelastic flow[26-27, 5]. It is suggested that for occurrence of instability and transition to turbulence, the critical condition is not completely dependent on the Reynolds number. However, the energy gradient function is a universal dimensionless number for characterizing the flow instability[5, 26-27].

## 2. Computational results with LES

The unsteady three-dimensional Navier-Stokes equation for incompressible fluid is the governing equation for laminar flow, transitional flow, and turbulent flow. The numerical methods to simulate these flows include the direct numerical simulation (DNS) and the large eddy simulation (LES) methods. The LES method has been used to simulate various wall bounded flows and wake flow after blunt bodies[28-31]. In the past twenty years or so, the LES method has been employed to compute the three-dimensional turbulent flows in Taylor-Couette flows as well as other wall-bounded flows. It has been shown that LES method is able to achieve almost identical numerical result with DNS[32-34]. Therefore, the LES method is employed in the present study. In addition, the identification methods of vortices developed in recent years enable us to track the vortex flow in transitional and turbulence[35-36]. The numerical method has been validated with the experimental results for the Taylor-Couette flow in Wereley and Lueptow[37].

The radius ratio of the inner cylinder to the outer cylinder is $R_1/R_2=0.83$, which corresponds to the inner cylinder and the outer cylinder of 43.4 mm and 52.3 mm. The length of the concentrated cylinders has a ratio of 46.6 to the gap width. The critical Reynolds number for the primary instability to produce the laminar Taylor cortex cell is $Re_{hc}=102$. The simulation result for transitional flow with this geometry is taken as the model "A" in the present study, as shown in Table 1.

Based on the model "A," the radius and the rotating speed of the inner cylinder are fixed and the radius of the outer cylinder is increased. Thus, the model "B" and the model "C" are obtained for different gap width as in Table 1.

Table 1 Parameters for the Taylor-Couette flow with three different gap widths

| Case | $R_1$(mm) | $R_2$(mm) | $h$(mm) | $L$(mm) | $\omega_1$(rad/s) | $Re_h$ |
|------|-----------|-----------|---------|---------|-------------------|--------|
| A    | 43.4      | 52.3      | 8.9     | 262.55  | 13.0              | 1600   |
| B    | 43.4      | 87.9      | 44.5    | 262.55  | 13.0              | 8000   |
| C    | 43.4      | 488.4     | 445     | 1050.2  | 13.0              | 80000  |

For the large eddy simulation (LES) in this study, the filtered unsteady Navier-Stokes equation and the wall-adapting local eddy viscosity(WALE) model for small scale turbulence are employed. The pressure-based finite volume method to solve the incompressible Navier-Stokes equations is employed to perform the simulations. The convected term in the governing equations is discretized with the second-order upwinding scheme and the diffusion term is approximated with the second central-difference scheme.

The resolution can be judged by the maximum of $y^+$ of the first layer of grid near the wall, where $y^+ = \sqrt{\tau_w/\rho}$, $\tau_w$ is the shear stress at the wall and $\rho$ is the density. In the present study, the value of $y^+$ is 0.09, 0.43, and 0.27 for three cases, respectively. These resolution corresponds to grid numbers of 4,194,304, 4,014,080, 3,932,160 grid nodes, respectively.

### 3.1 Averaged velocity distribution

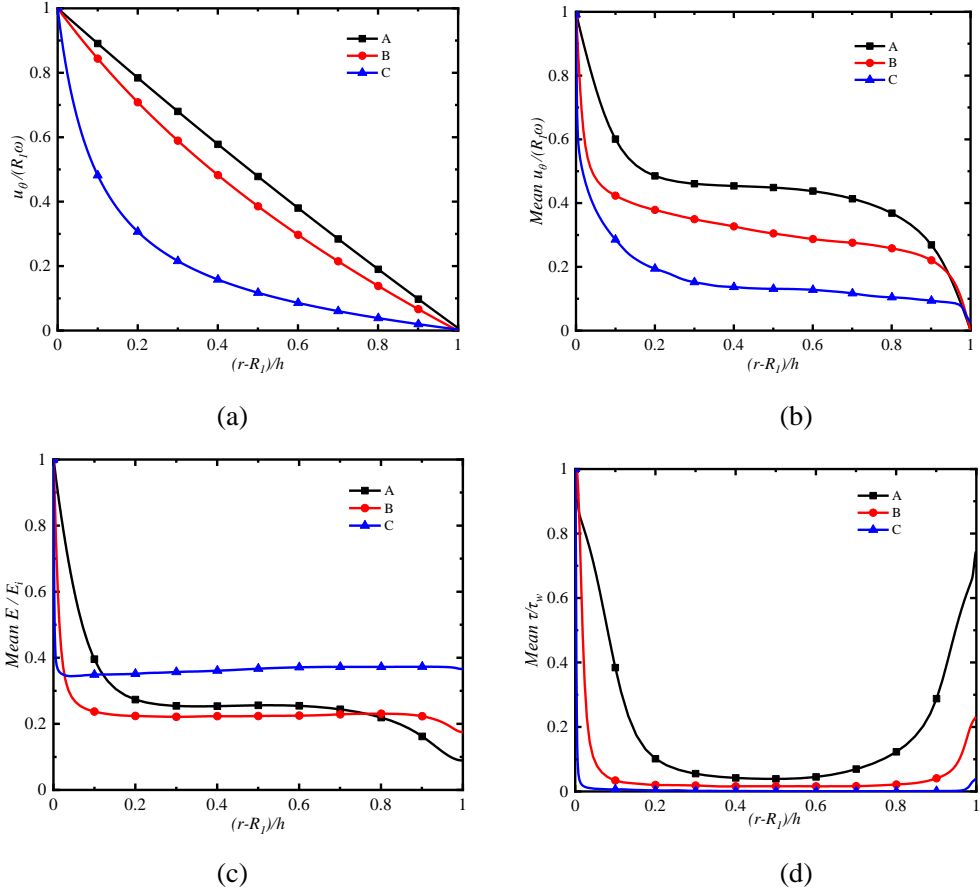

Fig.1 Distribution of the flow parameters along the radial direction for three different Reynolds numbers. (a) Velocity distribution in laminar flow; (b) Average velocity distribution in transitional flow. (c) Averaged total mechanical energy in transitional flow; (d) Averaged shear stress in transitional flow.

Figure 1(a) shows the tangential velocity distribution for laminar flow in the gap between the two rotating cylinders at three different Reynolds numbers, 1600, 8000, and 80000. It can be seen that when the gap width is narrow, the velocity distribution in the gap presents an approximate linear distribution. As the gap width gradually increases, the velocity distribution gradually approaches to the distribution of the free vortex flow. When the gap width is very large and the outer radius $R_2$ approaches infinity, the velocity distribution becomes almost a free vortex distribution. It has been demonstrated that the free vortex flow is the most stable based on the energy gradient theory [5].

Figure 1(b) shows the averaged tangential velocity distribution for transitional flow in the gap between two cylinders at three different Reynolds numbers. It shows that the tangential velocity distribution in transitional flow is distorted, and it becomes flat in the middle part of the gap as the increase of the Reynolds number. With the further increase of the gap width, the velocity distribution tends to that as in free vortex flow roughly. On the whole, the velocity distribution exhibits a similar trend with the variation of gap width as that in the laminar flow. As the gap width gradually increases, the averaged velocity distribution gradually approaches more likely to the distribution of free vortex. Among these three cases, the velocity distribution of the case C is more like a free vortex flow, and should be more stable. Therefore, the transition to turbulence should be delayed in case C.

Figures 1(c) and 1(d) show the distributions of the averaged total mechanical energy and the averaged shear stress of the transitional flow in the gap at three different Reynolds numbers. It can be seen from Fig.1(c) that when the gap width is narrow, the total mechanical energy within the gap has varied from the inner cylinder to the outer cylinder. As the gap width gradually increases, the total mechanical energy tends to be a constant. A constant mechanical energy is not able to amplify any disturbance.

In addition, it can be seen from Fig.1(d) that when the gap width is narrow (case A), the shear stress within the gap is higher. As the gap width increases, the shear stress gradually decreases. When the gap width is very large (case C), the shear stress in the gap becomes very small. With such low level of shear stress, velocity fluctuation is difficult to be transported along the radial direction.

### 3.2 Fluctuation of the circumferential velocity and the turbulent kinetic energy

Figure 2 shows the distributions of the fluctuation of the circumferential velocity and the turbulent kinetic energy in the gap computed using the LES method at three different Reynolds numbers. It can be seen from Fig.2(a) that when the gap width is narrow, the turbulent kinetic energy in the gap is large within the whole gap. As the gap width gradually increases, the turbulent kinetic energy in the gap gradually decreases. It can also be seen that the high turbulent kinetic energy is concentrated near the inner cylinder. When the gap width is large, the turbulent kinetic energy becomes very small. Therefore, the result in case C indicates that the flow close to the free vortex flow is of high stability, and the pulsations of the flow can be absorbed, making it less likely to generate turbulence. It can be seen that as the $Re_h$ increases, the turbulent kinetic energy decreases and the amplitude of the turbulent pulsation is reduced (Fig.2(b)).

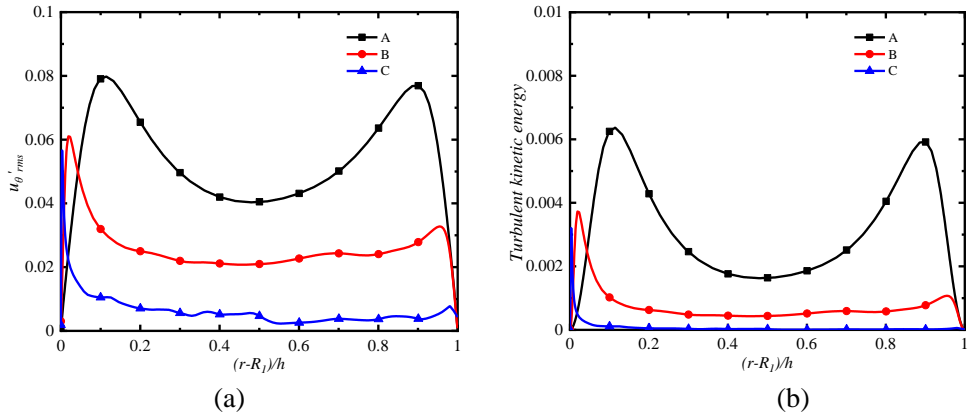

Fig.2 Distributions of the circumferential velocity fluctuations and the turbulent kinetic energy distribution in transitional flow at three different Reynolds numbers. (a) Radial distribution of the circumferential velocity fluctuations. (b) Radial distribution of the turbulent kinetic energy.

### 3.3 Instantaneous velocity distribution with time

Figure 3 shows the instantaneous velocity distribution in the transitional flow in the gap between two cylinders computed using the LES method at three different Reynolds numbers. It was described that the generation of the turbulent fluctuation depends on the generation of spikes in streamwise velocity [5, 22-24]. It can be seen from Fig.3 that as the gap width increases, the amplitude of the "spike" of velocity pulsation decreases, which delays the transition to turbulence. In other words, the increase of the $Re_h$ delays the generation of spikes, which postpones the generation of turbulence.

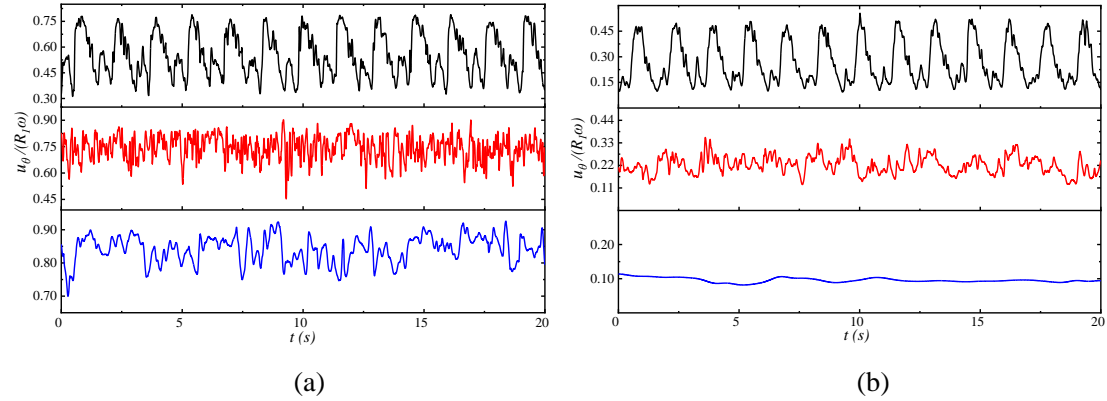

(a)            (b)

Fig.3 Instantaneous circumferential velocity at two positions in the gap between the two cylinders. (a) At the position (near the inner cylinder) with the maximum turbulent kinetic energy, (A) $(r-R_1)/h=0.12$; (B) $(r-R_1)/h=0.01$; (C) $(r-R_1)/h=0.0011$. (b) At the position (near the outer cylinder) with the high level of turbulent kinetic energy, at $(r-R_1)/h=0.9$. (In the two pictures, from top to bottom, A, B, and C).

### 3.4 Spectrum of the turbulent kinetic energy

The turbulent kinetic energy can be calculated by the following equation,

$$E_d = \frac{1}{2}\left(u'^2+v'^2+w'^2\right). \tag{6}$$

where $u'$, $v'$, $w'$, is the fluctuation velocity in the tangential, radial, and axial directions, respectively.

Kolmogorov[38] obtained a law of turbulent kinetic energy for isotropic homogeneous turbulence at high Reynolds number,

$$E_d = C_1\varepsilon^{2/3}k^{-5/3} \tag{7}$$

Here, $C_1$ is a constant, $\varepsilon$ is the dissipation rate, and $k$ is the wave number. This equation has achieved good agreement with a lot of simulations and experiments at high Reynolds number conditions. The Eq.(7) is generally called the K41 law of turbulent energy spectrum.

Figure 4 shows the variation of the turbulent kinetic energy with the disturbance frequency at the position of $(r-R_1)/h=0.5$ for three different Reynolds numbers. The high frequency corresponds to the larger wave number. It can be seen from Fig.4 that the turbulence kinetic energy in case A and case B has higher level of values, while the turbulent kinetic energy in case C is very small. This is owing to that the flow within the large gap width (case C) is close to a free vortex, which is very stable and is not prone to turbulence.

For case A and B, the energy spectrum only indicate a short frequency range which is near the K41 law, while for case C, the energy spectrum does not display such behavior of turbulence.

The reason why the curves in case A and B in Fig.4 does not follow the $-5/3$ slope can be explained as follow. The scaling law of turbulence energy spectrum by the K41 theory follows a power of $-5/3$, which corresponds to an isotropic homogeneous turbulence under high Reynolds number conditions. In the present study, (1) the Reynolds number is not high enough for producing isotropic homogeneous turbulence; (2) for Taylor-Couette flow, due to the fact that the flow in the gap in Taylor-Couette flow relies on energy input by the cylinder rotation, the flow is strongly anisotropic due to the work done, which results in a turbulence energy spectrum deviating from the $-5/3$ law.

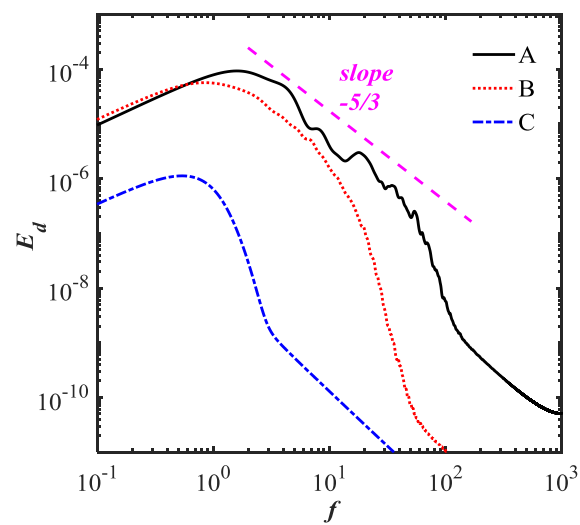

Fig.4 Spectrum of turbulent kinetic energy at the middle cylindrical surface of the gap for three different Reynolds numbers ($(r-R_1)/h=0.5$).

## 3.5 Distribution of the energy gradient function $K$

Figure 5 shows the variation of the energy gradient function $K$ along the gap width at three different Reynolds numbers, which are obtained at a given time moment. It can be seen from Fig.5 that the maximum of $K$ in the gap gradually decreases from case A to case C as a whole, indicating that the flow is becoming more stable. This is consistent with the tendency of the flow approaching a free vortex flow as the gap width is increased. As such, turbulence is less possibly generated as $Re_h$ increases to a very large value. This indicates that turbulent transition depends on the maximum of the energy gradient function $K$. In these cases, the Reynolds number $Re_h$ alone is not able to characterize the flow behavior, and the radius ratio should be employed together to express the behavior of flow stability in Taylor-Couette flows.

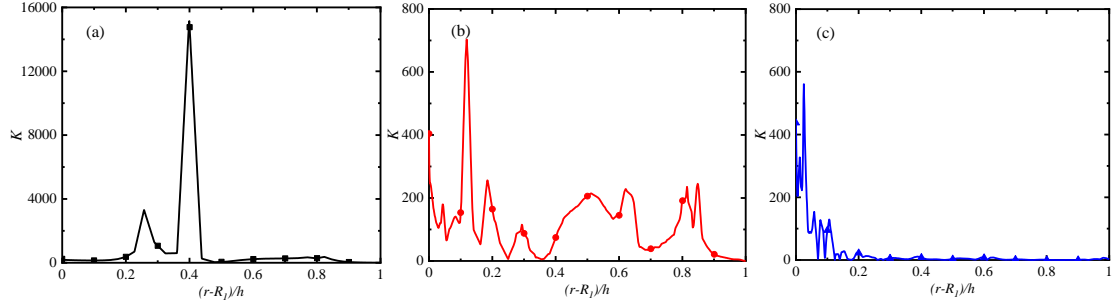

Fig.5 Instantaneous distribution of the energy gradient function $K$ along the radial direction at three different Reynolds numbers.

It can be found in Fig.5(a) that there is high peak of $K$ at the middle of the cylinder gap, nearly at $(r-R_1)/h=0.4$. This is caused by the fact that the value of the denominator in Eq.(5), $V\,\partial E/\partial s+\phi$, is nearly zero at the core area of the gap, as shown in Table 2. In the core area of the cylinder gap, the value of $V\,\partial E/\partial s$ is very small due to the approximate axisymmetrical flow, and the value of $\phi$ is nearly zero owing to zero shear stress. In Fig.5(b), the high peak of the $K$ appears near the inner cylinder at $(r-R_1)/h=0.1$. In Fig.5(c), the high peak of the $K$ appears near the inner cylinder at $(r-R_1)/h=0.03$. In this latter case, the value of the numerator in Eq.(5) is very small, which leads to the value of $K$ is low over the whole gap width. When it is approaching the inner cylinder, the value of $K$ becomes large. This is caused by the difference of the flow from the ideal free vortex flow.

Table 2 Values of various terms of Eq.(5) at the peak of function $K$ shown in Fig.5

| Cases | $K_{max}$ location | $V\,\partial E/\partial n$ | $V\,\partial E/\partial s$ | $\phi$ | $V\,\partial E/\partial s+\phi$ | $K$ |
|---|---|---|---|---|---|---|
| A | Gap middle | 227.13 | -1.7 | 1.685 | 0.015 | 15142 |
| B | Near inner wall | 2491.43 | -16.896 | 20.44 | 3.544 | 703 |
| C | Near inner wall | 686.10 | -0.757 | 1.98 | 1.223 | 561 |

**3.6 Contours of tangential velocity variation in the gap for the three cases**

The iso-surfaces of the tangential velocity for the three cases of gap widths are shown in Fig.6. It can be seen that obvious deep dimples are formed on the flow velocity iso-surface due to the appearance of the negative velocity spikes for case A. For case B, these dimples become shallow due to that the flow is a little more stable. For case C, It is seen that there is almost no dimples on the iso-surface of the flow velocity. These pictures show that the flow in case C is the most stable. The physical mechanism behind these phenomena is that the case C is nearer the free vortex flow than the other two cases.

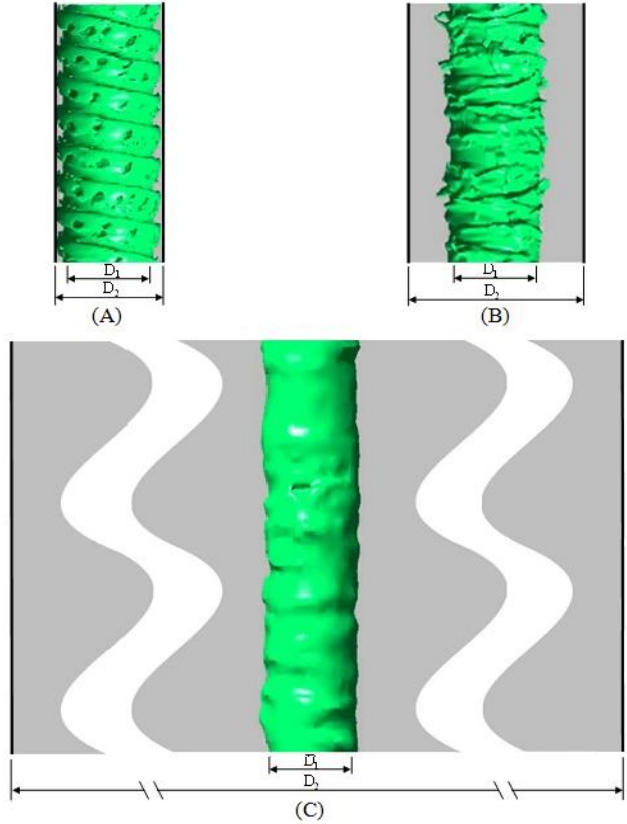

Fig.6 Iso-surface of dimensionless tangential velocity $u_\theta/(\omega R_1)$ for the three cases of gap width.
(A) $u_\theta/(\omega R_1)=0.45$; (B) $u_\theta/(\omega R_1)=0.45$; (C) $u_\theta/(\omega R_1)=0.45$.

**4 Conclusions**

In this study, the flow stability and turbulent transition in Taylor-Couette flow are studied by numerical simulation with LES, where the inner cylinder is rotating and the outer cylinder is fixed. The rotating speed of the inner cylinder is kept constant, and the radius of the outer cylinder is increased. Thus, the shear Reynolds number based on the gap width is varied. The simulation results show that for Taylor-Couette flow, with the gap width increasing, the flow becomes more stable, and the turbulent transition is delayed. The reason is that as the gap width increases, the proportion of the free vortex occupied in the flow field increases, enhancing stability. The main conclusions are summarized as follows.

(1) For Taylor-Couette flow, with the increase of the gap width, the fluctuation in tangential velocity decreases, and the turbulent transition is delayed.

(2) With the increase of the gap width, the flow in the cylinder gap gradually tends to free vortex flow, and thus the flow becomes more stable.

(3) As the gap width varies, the Reynolds number $Re_h$ alone is not able to characterize the flow behavior, and the radius ratio of cylinders should be taken into account.

(4) Turbulent transition depends on the magnitude of the energy gradient function $K$, which is able to characterize the flow stability in Taylor-Couette flows.

**Acknowledgement**

The authors thank Dr. Fu-jun Wang for the helps and discussions.

**References**

[1] Reynolds O. An experimental investigation of the circumstances which determine whether the motion of water shall be direct or sinuous, and of the law of resistance in parallel channels [J]. *Philosophical Transactions of the Royal Society of London. Series A*, 1883, 174: 935-982.

[2] Schlichting H., Gersten K. Boundary layer theory [M]. 9th Edition, Berlin, Germany: *Springer*, 2017.

[3] Yaglom A. M. Hydrodynamic instability and transition to turbulence [M]. Frisch, U.(Ed.), Dordrecht, The Netherlands: *Springer*, 2012.

[4] Lumley J., Yaglom A. M. A century of turbulence [J]. *Flow, Turbulence and Combustion*, 2001, 66: 241-286.

[5] Dou H. S. Origin of turbulence-energy gradient theory [M]. Singapore: *Springer*, 2022.

[6] Taylor G. I. Stability of a viscous liquid contained between two rotating cylinders [J]. *Philosophical Transactions of the Royal Society of London. Series A*, 223, 289-343.

[7] Coles D. Transition in circular Couette flow [J]. *Journal of Fluid Mechanics*, 1965, 21: 385-425.

[8] Andereck, C. D., Liu S.S., Swinney H. L. Flow regimes in a circular Couette system with independently rotating cylinders [J]. *Journal of Fluid Mechanics*, 1986, 164, 155-183.

[9] Coughlin K., Marcus P. S. Turbulent Bursts in Couette-Taylor Flow [J]. *Physical Review Letters*, 1996, 77 (11), 2214-2217.

[10] Prigent A., Grégoire G., Chaté H. et al. Large-scale finite-wavelength modulation within turbulent shear flows [J]. *Physical Review Letters*, 2002, 89(1): 014501.

[10] Drazin P. G., Reid W. H. Hydrodynamic stability [M]. 2nd Edition, Cambridge, UK: *Cambridge University Press*, 2004.

[11] Feldmann D., Borrero-Echeverry D., Burin M. J. et al. Routes to turbulence in Taylor–Couette flow [J]. *Philosophical Transactions of the Royal Society A*, 2023, 381, 20220114.

[11] Merbold M. H. Hamede A. Froitzheim C. E. Flow regimes in a very wide-gap Taylor–Couette flow with counter-rotating cylinders [J]. *Philosophical Transactions of the Royal Society A*, 381, 20220113.

[12] Shi L., Hof B., Rampp M. et al. Hydrodynamic turbulence in quasi-Keplerian rotating flows [J]. *Physics of Fluids*, 2017, 29, 044107.

[13] Razzak M. A., Khoo B. C., Lua K. B. Numerical study on wide gap Taylor Couette flow with flow transition [J], *Physics of Fluids*, 2019, 31, 113606.

[14] Ji H., Goodman J. Taylor-Couette flow for astrophysical purposes [J]. *Philosophical Transactions of the Royal Society A*, 2023, 381, 20220119.

[15] Crowley C. J., Pughe-Sanford J. L., Toler W. et al. Turbulence tracks recurrent solutions [J]. *Proceedings of the National Academy of Sciences of the United States of America*, 2022, 119(34), e2120665119.

[16] Bilson M., Bremhorst K. Direct numerical simulation of turbulent Taylor–Couette flow [J]. *Journal of Fluid Mechanics*, 2007, 579, 227–270.

[17] Froitzheim A., Ezeta R., Huisman S. G. et al. Statistics, plumes and azimuthally travelling waves in ultimate Taylor–Couette turbulent vortices [J]. *Journal of Fluid Mechanics*, 2019, 876, 733–765.

[18] Dubrulle B., Dauchot O., Daviaud F. et al. Stability turbulent transport in Taylor-Couette flow from analysis of experimental data [J]. *Physics of Fluids*, 2005, 17, 095103.

[19] Dou H. S., Khoo B. C., Yeo K. S. Instability of Taylor-Couette flow between concentric rotating cylinders [J]. *International Journal of Thermal Sciences*, 2008, 47, 1422-1435.

[20] Dou H. S. Singularity of Navier-Stokes equations leading to turbulence [J]. *Advances in Applied Mathematics and Mechanics*, 2021, 13(3), 527-553.

[21] Dou H. S. No existence and smoothness of solution of the Navier-Stokes equation [J]. *Entropy*, 2022, 24, 339.

[22] Niu L., Dou H. S., Zhou C. et al. Turbulence generation in the transitional wake flow behind a sphere [J]. *Physics of Fluids*, 2024, 36, 034127.

[23] Berghout P., Dingemans R.J., Zhu X. et al. Direct numerical simulations of spiral Taylor–Couette turbulence [J]. *Journal of Fluid Mechanics*, 2020, 887, A18.

[24] Dou, H. S., Phan-Thien N. Viscoelastic flows around a confined cylinder: Instability and velocity inflection [J]. *Chemical Engineering Science*, 2007, 62(15), 3909-3929.

[25] Dou H. S., Phan-Thien N. An instability criterion for viscoelastic flow past a confined cylinder [J]. *Korea and Australia Rheology*, 2008, 20, 15-26.

[26] Xu C., Chen L., Lu X. Large-eddy and detached-eddy simulations of the separated flow around a circular cylinder [J]. *Journal of Hydrodynamics*, 2007, 19(5), 559–563.

[27] Deng Y., Chong K., Li Y. et al. Large-eddy simulation of turbulent boundary layer flow over multiple hills [J]. *Journal of Hydrodynamics*, 2023, 35, 746–756.

[28] Tan L., Zhu B., Wang, Y. et al. Turbulent flow simulation using large eddy simulation combined with characteristic-based split scheme [J]. *Computers and Fluids*, 2014, 94, 161-172.

[29] Wei Q., Chen H., Ma Z. An hybrid RANS/LES model for simulation of complex turbulent flow [J]. *Journal of Hydrodynamics*, 2016, 28(5), 811–820.

[30] Schlatter P., Stolz S., Kleiser L. Large-eddy simulation of spatial transition in plane channel flow [J]. *Journal of Turbulence*, 2006, 7(1), 1-24.

[31] Poncet S., Viazzo S., Oguic R. Large eddy simulations of Taylor-Couette-Poiseuille flows in a narrow-gap system [J]. *Physics of Fluids*, 2014, 26, 105108.

[32] Razzak M. A., Khoo B. C., Lua K. B. Numerical study of Taylor-Couette flow with longitudinal corrugated surface [J]. *Physics of Fluids*, 2020, 32, 053606.

[33] Wang Y., Gao Y., Liu J. et al. Explicit formula for the Liutex vector and physical meaning of vorticity based on the Liutex-Shear decomposition [J]. *Journal of Hydrodynamics*, 2019, 31(3), 464-474.

[34] Dong X., Cai X., Dong Y. et al. POD analysis on vortical structures in MVG wake by Liutex core line identification [J]. *Journal of Hydrodynamics*, 2020, 32(3), 497-509.

[35] Wereley S. T., Lueptow R. M. Velocity field for Taylor–Couette flow with an axial flow [J]. *Physics of Fluids*, 1999, 11(12), 3637-3649.

[36] Kolmogorov A. N. The local structure of turbulence in incompressible viscous fluid for very large Reynolds numbers [J]. *C R Acad Sci URSS*, 1941, 30, 301-305.